\documentstyle[12pt,epsfig]{article}

\begin{document}

\baselineskip=18.6pt plus 0.2pt minus 0.1pt

\begin{titlepage}
\title{
\begin{flushright}
 {\normalsize GNPHE/03-05}\\
\mbox{}
\end{flushright}  
{\bf On $N=1$ gauge models from geometric engineering in M-theory}
}
\author{A. Belhaj$^{1,2,}$\thanks{{\tt belhaj@crm.umontreal.ca}} ,\ \
 L.B. Drissi$^{1}$,\ \
J. Rasmussen$^{2,3,}$\thanks{{\tt rasmusse@crm.umontreal.ca}}
\\[8pt]
{\it \small $^1$ National Grouping of High Energy Physics, GNPHE, 
{\em and}}\\  
{\it \small Lab/UFR High Energy Physics, Department of Physics}\\
{\it \small Faculty of Sciences, Rabat, Morocco}
\\[8pt]
{\it \small $^2$  Department of  Mathematics  and Statistics, 
 Concordia University}\\
{\it \small  Montr\'eal, PQ,  Canada  H4B 1R6}
\\[8pt]
{\it\small $^3$ Centre de Recherches Math\'ematiques, Universit\'e de 
Montr\'eal} \\
{\it\small C.P. 6128, succursale centre-ville, Montr\'eal, PQ, Canada H3C 3J7}
}   
\maketitle \thispagestyle{empty}
\begin{abstract}
We study geometric engineering of four-dimensional  
$N=1$ gauge  models from M-theory on a seven-dimensional manifold 
with $G_2$ holonomy. The manifold is constructed as a K3 fibration over a
three-dimensional base space with $ADE$ geometry. The resulting
gauge theory is discussed in the realm of $(p,q)$ webs.
We discuss how the anomaly cancellation condition translates into 
a condition on the associated affine $ADE$ Lie algebras. 
\end{abstract}
  
\end{titlepage}
  
  \def\be{\begin{equation}}
  \def\ee{\end{equation}}
  \def\bea{\begin{eqnarray}}
  \def\eea{\end{eqnarray}}
  \def\nn{\nonumber}
  \def\l{\lambda}
  \def\t{\times}
  \def\[{\bigl[}
  \def\]{\bigr]}
  \def\({\bigl(}
  \def\){\bigr)}
  \def\p{\partial}
  \def\o{\over}
  \def\ta{\tau}
  \def\cm{\cal M}
  \def\R{\bf R}
  \def\b{\beta}
  \def\a{\alpha}
\newpage

\section{Introduction}

Type IIA superstring compactification on a K3 space with
$A_{n-1}$ singularities 
develops an enhanced $SU(n)$ gauge symmetry in six dimensions \cite{KV}. 
This can be further compactified on a real two-dimensional base space,
resulting in a four-dimensional $N=2$ quantum field theory (QFT$_4$).
The gauge fields and (massless) matter fields are described by
wrappings of D-branes around cycles in the
Calabi-Yau threefold defined by the K3 fibrations.
The gauge group and matter content of the QFT$_4$ thus depend on
the non-trivial base geometry.
This program is referred to as geometric engineering of $N=2$ 
QFT$_4$. These models, which give exact 
results for the moduli space of the Coulomb branch, may be represented by 
quiver diagrams similar to Dynkin diagrams of ordinary or affine
Lie algebras \cite{KLMVW,KatzVafa,KKV,BJPSV,KMV,Mayr,BFS}. 
Let the base space consist of a collection of intersecting two-spheres,
each of which will give rise to an $SU$ gauge-group factor. To each of 
these is associated a quiver or Dynkin diagram node, and for each pair
of groups with matter in bi-fundamental representations, 
the two corresponding nodes are connected by a line.

One should distinguish three cases: models with negative beta function, models
with conformal invariance, requiring the vanishing
of the beta function, and models with positive beta function.   
The first models are represented by ordinary 
$ADE$ Dynkin diagrams with only bi-fundamental matter.
A K3 fiber with $A_{n_i-1}$ singularities generates an $SU(n_i)$ 
gauge-group factor, resulting in the gauge group
\be
 G\ =\ \bigotimes_{i=1}^mSU(n_{i}),
\ee 
where $m$ is the number of intersecting two-spheres.
The second models are classified by two categories according to the 
type of intersecting geometry of the base space: 
(i) $N=2$ superconformal field theory in four dimensions (SCFT$_{4}$), 
based on \textit{finite} $ADE$ Lie algebra
singularities. It has gauge group similar to the first class
but matter in both fundamental $({n}_{i})$ and bi-fundamental 
$({n}_{i},{\bar {n}}_{j})$ representations of $G$. 
(ii) $N=2$ SCFT$_{4}$ with gauge group
\be
 G\ =\ \bigotimes_{i=1}^mSU(s_{i}n)
\label{SUsin}
\ee 
and bi-fundamental matter only. Here $n$ is an unspecified positive
integer. This second category of
scale invariant field models is classified by \textit{affine} $ADE$ Lie
algebras. The positive integers $s_{i}$ appearing in $G$ are the 
Coxeter labels (sometimes referred to as marks or Dynkin numbers); 
they form a positive-definite integer vector 
$\left(s_{j}\right)$ satisfying 
\be
 \sum_{j=1}^m{K}_{ij}s_{j}\ =\ 0.
\label{K}
\ee
$K$ is the Cartan matrix of the affine Lie algebra, and $m=r+1$ with
$r$ being the rank of the underlying Lie algebra.
Models with positive beta function are discussed in \cite{abs}. 
 
Quite recently, four-dimensional quiver gauge models preserving only 
four supercharges have attracted a lot of attention. Work has been done 
using  either type II superstrings on Calabi-Yau threefolds 
\cite{CSU,HI,FHHI,FH,Uranga,CIM}, or M-theory on $G_2$ manifolds 
\cite{He,Be}. The aim of this work is to contribute to this program.  
We study M-theory on seven-dimensional manifolds with $G_2$ holonomy group. 
They are realized explicitly as K3 fibrations over a three-dimensional base
space. First, though, we review the geometric engineering of $N=2$ QFT$_4$
in type IIA superstring theory. We then extend this to geometric
engineering of $N=1$ QFT$_4$ in M-theory, where particular
emphasis is put on the three-dimensional $ADE$ base geometry.
Comparing our construction to the related D6-brane scenario described
using the method of $(p,q)$ webs, we determine the gauge group
and matter content of our $N=1$ QFT$_4$.

\section{Geometric engineering of $N=2$ gauge theory}

A well-known way to get supersymmetric quantum field theory (QFT)  
from superstrings, M- or F-theory, is to consider the compactification on  
a {\em singular} manifold $X$ with K3 fibration over a base space 
$B$. The gauge group $G$ and matter content 
of the QFT are defined by the singularities of the fiber and 
the non-trivial geometry of the base space. 
In particular, the gauge coupling $g$ is proportional to the inverse of 
the square root of the volume of the base: $V(B)=g^{-2}$.
The complete set of physical parameters of the QFT is related to 
the moduli space of the manifold. 
Moreover, several exact results for the Coulomb branch of  
$N=2$ QFT$_4$ embedded in type IIA superstring theory are obtained naturally
by local mirror symmetry of Calabi-Yau threefolds $CY^3$.  
The latter are realized as a K3 fibration with $ADE$ singularity over
a projective $P^1$ complex curve 
or a collection of intersecting $P^1$ curves.

Let us briefly outline the main steps in obtaining
$N=2$ QFT$_4$ from type IIA superstring theory
on $CY^3$. The idea is subsequently illustrated by a simple example.  
Depending on the sought gauge group, one first specifies the  
$ADE$ singularity of the K3 fiber. Then one considers 
the limit in which the volume $V(B)$ of the base of $CY^3$ is 
very large so that gravitational effects may be ignored. This allows
one to consider locally (i.e., near the singularities)
the K3 fibers as non-compact ALE spaces
with the same singularities. Finally, one examines 
the propagation of the type IIA superstrings on $CY^3$ in the 
presence of D2-branes wrapping the vanishing two-cycles in the ALE
spaces. This enables one to make conclusions about the gauge group and
matter content.

As an illustration, we now consider the situation where the K3 fiber has 
an $A_1$ singularity. 
In the vicinity of the singularity, such a fiber may be viewed 
as an ALE space with $A_1$ singularity and described by 
\be
 z_1^2+z_2^2+z_3^2=0.
\ee
Using a simple change of variables, this is equivalent to
\be  
 xy=z^2,
\ee
where $x$, $y$ and $z$ are complex variables.
This geometry has a nice physical realization as a two-dimensional $N=2$ 
linear sigma model with $U(1)$ gauge symmetry. To see this, consider the 
three chiral  fields $\phi_i$ with charges $(1,-2,1)$. The coordinates of the
ALE space can then be expressed in terms of $U(1)$ gauge invariants:
\bea
 x&=&\phi_1^2\phi_2,\nn\\
 y&=&\phi_3^2\phi_2,\\
 z&=&\phi_1\phi_2 \phi_3.\nn
\eea
This is related to the so-called  
$D$-term in the bosonic potential $U(\phi_1, \phi_2,\phi_3)$ 
in supersymmetric theories with four supercharges:
\be
 U(\phi_1, \phi_2,\phi_3)=(\phi_1 \phi_1+\phi_3\phi_3-2\phi_2\phi_2-R)^2.
\label{U}
\ee
In this equation, $R$ is the coupling parameter of the 
$U(1)$ Fayet-Iliopoulos (FI) term one may introduce in the Lagrangian model.
In the superfield language, the action then reads
\be
 S(\Phi,V)= \int d^2xd^4\theta \bar\Phi e^V \Phi -R\int d^2xd^4\theta V,
\label{S}
\ee
where $\Phi$ and $V$ are the chiral and gauge superfields, respectively.
{}From this, one sees that the $U(1)$ Cartan subgroup of 
the $SU(2)$ symmetry of the singularity of K3 carries the gauge 
symmetry of the $N=2$ supersymmetric  linear sigma model. The presence 
of the FI term resolves 
the singularity of the potential $U(\phi_1, \phi_2,\phi_3)$. 
Geometrically, this corresponds to replacing the singular point $x=y=z=0$ 
by a $P^1$ curve parameterized by a new variable $ x'$ defined 
as $x'={x/z}$. In the new local coordinates  $(x',y,z)$, the 
equation of the original $A_1$ singularity becomes
\be
  x' y=z
\ee
thus removing the singularity (which is now ``hidden'' in the
singular transformation $x'=x/z$). In the field theory language, 
this corresponds to a positive value of the FI coupling $R$.

The next step is to consider the propagation of type IIA superstrings 
in this background. In this case, D2-branes wrapping around the blown-up 
$P^1$ curves give rise to a pair of vector particles, $W^{\pm}$, one for
each of the two possible orientations of the wrappings. The particles have 
masses proportional to the volume of the blown-up (real) two-sphere
described by $x'$. 
$W^{\pm}$ are charged under the $U(1)$ field $Z$ obtained 
by decomposing the type IIA superstring three-form in terms of the 
harmonic form on the two-sphere and a one-form not in the K3  fiber. 
In the limit where the blown-up two-sphere shrinks to a point, 
the three vector particles become massless, and they form an $SU(2)$ 
adjoint representation. 
We thus obtain an $N=2$ $SU(2)$ gauge symmetry in six dimensions. 
A further compactification on the base $B_2$ (here chosen as a real 
two-sphere), gives $N=2$ pure $SU(2)$ Yang-Mills theory in four dimensions.

The geometric $SU(2)$ gauge theory analysis is 
extended straightforwardly to higher $SU(n)$ gauge groups and products thereof.
Extensions to general simply-laced $ADE$ gauge groups 
are obtained using more general Calabi-Yau threefolds \cite{KMV,Mayr}.
Up to some pertinent details, non-simply-laced $BCFG$ groups 
may also be constructed \cite {BFS,BS}, see also 
\cite{BPS,BKMT}. The incorporation of matter was alluded to in the 
Introduction, and is governed by a non-trivial base space geometry of the 
Calabi-Yau threefold. For example, consider 
a two-dimensional locus with an $SU(n)$ singularity and another 
locus with an $SU(m)$ singularity. If they intersect or meet at a point, the
(mixed) wrapped two-cycles give rise to bi-fundamental
matter of the $SU(n) \otimes SU(m)$ gauge group in four dimensions.
This extends readily to multiple intersecting loci.
It was also indicated in the Introduction how the associated toric quiver 
diagrams of $N=2$ QFT$_4$'s are similar
to the usual Dynkin diagrams of ordinary and affine Lie algebras.

\section{Geometric engineering of $N=1$ gauge theory}

Here we discuss a straightforward way of elevating the geometric 
engineering of $N=2$ gauge models in string theory to a similar 
construction of $N=1$ gauge models in M-theory. 
Alternative studies of four-dimensional $N=1$ gauge models in the
framework of F- or M-theory may be found in refs. \cite{V,acha,FW}.

\subsection{Outline of method}

The method we will be using here is quite similar to the   
geometric engineering of $N=2$ QFT$_4$ discussed above. 
Our starting point is a local description of M-theory 
compactification on a $G_2$ manifold with K3 fibration over  
a real three-dimensional base space $B_3$. In this way, the 
scenario of type IIA superstring theory in six dimensions appears in seven 
dimensions in M-theory. The D2-branes are 
replaced by M2-branes, and we obtain an $N=2$ gauge 
theory in seven dimensions. If we compactify this seven-dimensional theory   
on $T^3$ to four dimensions, we obtain an $N=4$ gauge theory, while
a compactification on $S^2\times S^1$ leads to a system 
with eight supercharges (corresponding to $N=2$) \cite{acha}.

We are, however, interested in models with only four supercharges in four 
dimensions, i.e., $N=1$ QFT$_4$. 
In order to reach such a model we need to compactify on a manifold 
whose three-dimensional base preserves $1/4$ of the remaining 
16 supercharges. The seven-dimensional K3 fibration must also 
have vanishing first Betti number, $b_1=0$, to meet the requirement
of $G_2$ holonomy. An example of such a geometry has the 
three-dimensional sphere $S^3$ as base, in which case we obtain only 
{\em pure} $N=1$ Yang-Mills theory \cite{acha,FW}.
We shall, instead, consider (a collection of) non-trivial $S^2$
fibrations over a finite line segment. 

As in the $N=2$ case, the incorporation of   
matter may be achieved by introducing a non-trivial geometry in the 
base of the K3 fibration. This leads us to consider
a three-dimensional intersecting geometry to describe a product
gauge group with bi-fundamental matter.

\subsection{Construction of the base geometry}

Our construction of the non-trivial base geometry $B_3$ is quite simple
and motivated by recent work on Lagrangian submanifolds in Calabi-Yau 
manifolds \cite{GVW}. We shall consider the three-dimensional base space
as a two-dimensional fibration over a one-dimensional base, where 
the fiber and the base each preserves half of the 
seven-dimensional M-theory  supercharges. The entire base space $B_3$
is embedded in a three-dimensional complex Calabi-Yau manifold.
As discussed below, this ensures that the elements of the intersection matrix
of a collection of such base space constituents are integers.
The Calabi-Yau manifolds are realized explicitly as a family of (deformed)
$ADE$ singular K3 surfaces over the complex plane. 
This geometry extends the family of elliptic curves on the complex plane 
used in F-theory compactifications down to eight dimensions \cite{Vafa96}.
In this way, non-trivial three-cycles, satisfying the 
constraints of the $G_2$ base geometry, constitute an 
intersecting geometry of $ADE$ K3 surfaces fibered over a line segment  
in the complex plane. The ``$ADE$ part'' corresponds to a two-dimensional
geometry obtained by resolving the $ADE$ singularities of the K3 surfaces. 
As already mentioned, the $ADE$ geometry is then either an (i)
ordinary (finite Lie algebraic) $ADE$ geometry, or an (ii) 
affine (infinite Lie algebraic) $ADE$ geometry.

We start by examining the model with an ordinary 
$A_1$ singularity in the K3 fiber. This is subsequently extended
to more general $ADE$ geometries. The deformed $A_1$ geometry 
is given by 
\be
 z_1^2+z_2^2+z_3^2= \mu 
\label{mu}
\ee 
where $\mu$ is a complex parameter. The real part of this parameter, 
being the radius squared of the sphere, is related to the FI terms 
(\ref{U}) and (\ref{S}), while the imaginary part can be identified 
with the NS-NS $B$-field in superstring compactifications. The $A_1$ 
threefolds may be obtained by varying the parameter $\mu$ over 
the complex plane parametrized by $w$:
\be
 z_1^2+z_2^2+z_3^2= \mu(w).
\ee 
The base space $B_3$ can then be viewed as a finite line segment with an 
$S^2$ fibration, where the radius $r$ of $S^2$ vanishes at the two interval 
end points, and at the end points only \cite{HI}. The latter requirement
ensures that no unwanted singularities are introduced.
One way to realize this geometry is 
\be 
 r\ \sim\ \sin x
\label{r}
\ee
where $x$ is a real variable parameterizing the interval $\[0,\pi\]$ 
in the $w$-plane. 

This construction may be extended to more general $ADE$
geometries where we have intersecting  spheres according to affine $ADE$
Dynkin diagrams. Here we consider a general model describing
a family of deformed elliptic $ADE$ singularities of K3
surfaces fibered over the complex plane, and their mirror partners in type II
superstring compactifications.
In type IIB superstring theory, the blow-up of these
singularities can be described in a two-dimensional $N=2$ linear sigma model 
with $U(1)^{\otimes(r+1)}$ gauge symmetry. It has $r+5$ chiral
multiplets $\phi_\ell$ with vector charges $q^i_\ell$, where $r$ is the
rank of the associated affine $ADE$ Lie algebra.
The $D$-terms corresponding to such classical vacua are given by
\be
 U=\sum\limits _{i=1}^{r+1}\sum \limits _{\ell=1}^{r+5}(q_\ell^i
  |\phi_\ell|^2-R_i)^2,
\ee 
where the FI terms are denoted $R_i$.
The $q^i_\ell$'s, being  the quantum charges of $\phi_\ell$ 
under the $U(1)^{\otimes(r+1)}$ gauge symmetry, are identified
with the intersection matrix (known as Mori vectors in toric geometry) 
of the non-compact and compact divisors of the elliptic $ADE$ K3 surfaces.
For the compact divisors\footnote{We shall be using a convention where the
first $r+1$ divisors are compact, i.e., $q^j_1,\ldots,q^j_{r+1}$ 
correspond to the compact divisors.} used in the resolution of the elliptic  
$ADE$ singularities, the intersection numbers are given by minus
the affine $ADE$ Cartan matrix $K$:
\be
  q_i^j=-K_{ij}, \quad i,j=1,\ldots,r+1.
\ee 
This link stems from a nice correspondence between the $ADE$ roots   
and the two-cycles, $C$, involved in the deformation of the elliptic 
singularities of the K3 surfaces \cite{KMV}. In this way, the cohomology 
class of the deformed elliptic curve, $T^2$, is represented by
 \be
 [T^2]=\sum_{i=1}^{r+1} s_i C_i,
\label{TC}
\ee
where $s_i$ are the Coxeter labels of the affine Lie algebra. 

The elliptic $ADE$ threefolds
may now be described using the toric geometry data of these threefolds
and local mirror symmetry \cite{Mayr,BFS,BS}. Applying this symmetry
in the K3 fiber, the elliptic $ADE$ type IIA mirror geometries, 
$ X^{\ast }$, may be described by an algebraic geometry equation of the form
\be
 P\left( X^{\ast }\right)=\sum_{\ell=1}^{r+5}a_{\ell} (w)y_{\ell}.
\label{210}
\ee
$a_{\ell}=a_{\ell}(w)$ are varying complex 
moduli\footnote{Note that only $r+1$ of these are physical.} in the $w$-plane,
while the $y_{\ell}$'s are complex variables satisfying
\be
  \prod\limits_{\ell=1}^{r+5}y_{\ell}^{q_{\ell}^{i}}=1, \quad
  \forall\; i=1,\ldots,r+1.
\label{mirror}
\ee
This defines the mirror elliptic $ADE$ K3  fiber. Eq. (\ref{mirror}) can be
solved by a quasi-homogeneous polynomial in some variables $(x,y,z,v)$, and 
eq. (\ref{210}) becomes
\be
 p_0(x,y,z;\lambda (w))+\sum_{i=1}^{r+1}
  p_i(x,y,z,v;a_i(w))=0.
\ee
Here $\left(x,y,z\right)$ are the homogeneous coordinates of
the weighted projective space ${WP}^{2}_{3,2,1}$, while the index on
$a_i$ refers to a (physical) compact divisor. Parameterized by the
complex structure modulus $\lambda$,
$p_0=0$ now describes a family of elliptic curves over the $w$-plane:
\be
 x^2+y^3+z^6+ \lambda (w) xyz=0.
\ee
The $v$-dependent term depends on the structure of the affine Lie
algebra. In particular, the powers of the complex variable $v$
appearing in the $ADE$ mirror geometries are given by the Coxeter labels
of the corresponding affine Lie algebra. The geometry of affine $A_{2}$, for 
instance, is given by
\be
 x^{2}+y^{3}+z^{6}+ \lambda (w) xyz+v\[ a_1(w)z^{3}+a_2(w) yz+a_3(w)x\]=0.
\ee

A similar analysis of the simple $A_1$ model discussed in eqs.
(\ref{mu})-(\ref{r}) could be done using the 
so-called mirror map in the Calabi-Yau moduli spaces \cite{KMV}. 
Indeed, mirror symmetry relates the complixified FI terms 
$T_j$ in type IIB geometry $X$ to the coordinates, $Z_j$, of the 
complex moduli space of type IIA geometry:
\be
 T_j=B_j+iV(C_j)\ \sim\ -i\ln(Z_j).
\label{map}
\ee
$V(C_j)$ is the volume of the two-cycle $C_j$, whereas $Z_j$ is given by 
\be
 Z_j=\prod\limits_{\ell=1}^{r+5}a_{\ell}^{q_{\ell}^{j}}.
\ee
For vanishing integrated NS-NS $B$-fields $B_j$, equation (\ref{map}) 
reduces to 
\be
 V(C_j)\ \sim\ -\ln(Z_j).
\label{rmap}
\ee
Under mirror symmetry, the constraints on the K\"ahler 
parameters of the two-cycles in type IIB geometry translate 
into conditions on the complex structures in type IIA geometry.
 
In the geometric setting, our choice of base geometry is given as a 
line segment fibered by intersecting 
spheres according to $ADE$ affine Dynkin diagrams.

\section{Gauge group and matter content}

Having specified the base geometry of the $G_2$ manifold, we  
will discuss the corresponding gauge theory of the compactified
M-theory. Our analysis here will be  based 
on a dual type IIA  superstring description. It turns out that, in addition 
to being dual to heterotic string compactification, M-theory 
on $G_2$ manifolds is also dual to type IIA string theory on Calabi-Yau 
threefolds in the presence of D6-branes wrapping Lagrangian sub-manifolds 
and filling the four-dimensional Minkowski space \cite{CSU}.  
In particular, a local description of M-theory near the $A_{n-1}$ singularity 
of K3 surfaces is equivalent to $n$ units of D6-branes \cite{LV}. 
Indeed, on the seven-dimensional world-volume of each D6-brane 
we have a $U(1)$ symmetry. When the $n$ D6-branes approach each other,
the gauge symmetry is enhanced from $U(1)^{\otimes n}$ to $U(n)$. 
An extra compactification of M-theory down to four-dimensional 
space-time is equivalent to wrapping D6-branes on the same geometry.
In this way, the physics content of the resulting D6-brane scenario can be 
computed explicitly using the method of the so-called $(p,q)$ webs 
\cite{HI,FHHI,FH,Uranga,CIM}. Using the results of this method,
we expect that the final gauge theory in four-dimensional Minkowski space
has gauge group
\be
 G\ =\ \bigotimes_{i=1}^m U(n_i).
\ee 
$m$ is the number of intersecting three-cycles in the base geometry.
The integers $n_i$ are specified by the anomaly cancellation 
condition. This means that they should form 
a null vector of the intersection matrix of the three-cycles:  
\be
 \sum_{j=1}^mI_{ij}n_j\ =\ 0.
\label{acc}
\ee 
In the infra-red limit the $U(1)$ factors decouple and one is left 
with the gauge group
\be
 G\ =\ \bigotimes_{i=1}^m SU(n_i).
\ee 

As already mentioned, the gauge group and matter content depend on the 
intersecting geometry in the three-dimensional base of the $G_2$ manifold. 
The intersection matrix of real $N$-dimensional spheres $S^N$ 
(constructed as blow-ups of singularities in Calabi-Yau spaces) 
is symmetric for $N$ even and anti-symmetric for $N$ odd. 
In more general geometries, the intersection matrix may be written as a 
linear combination of a symmetric and an anti-symmetric term.

Here we identify the base with a collection of
the three-cycles discussed in Section 3. We recall that they correspond 
to $S^2$ fibrations over a line segment. Thus, the intersection matrix of 
the three-cycles is equivalent to the intersection
matrix formed by two-spheres only. From the obvious similarity
with the $N=2$ scenario discussed in Section 2, we see that 
information on our intersecting geometry is naturally encoded
in the Cartan matrix, $K$, of an affine $ADE$ Lie algebra:
\be
 I_{ij}=-K_{ij}.
\ee
It follows that the anomaly cancellation condition is translated into 
a condition on the affine Lie algebra, thus solving (\ref{acc}) by
\be
 n_i=s_in.
\ee
Here $s_i$, $i=1,\dots,r+1$, are the Coxeter labels of the associated
affine Lie algebra of rank $r$, and $r+1$ is equal to the number of 
intersecting three-cycles. As in (\ref{SUsin}), $n$ is an unspecified positive
integer. The gauge group finally reads
\be
 G\ =\ \bigotimes_{i=1}^{r+1} SU(s_in).
\ee

\section{Conclusion}

We have studied the geometric engineering of $N=1$
four-dimensional models from M-theory on a manifold with
$G_2$ holonomy. This manifold is a K3 fibration over a 
three-dimensional base space with $ADE$ geometry. To that
end we have first constructed the base geometry as 
intersecting two-cycles over a line segment. This special base
has been embedded in a family of K3 surfaces over the complex
plane. Based on the relation between M-theory and D6-branes 
in type IIA superstrings, we have discussed the physics content of 
M-theory compactified on our $G_2$ manifold. In particular,
the resulting gauge theory has been discussed in the framework
of $(p,q)$ webs. In this scenario, the anomaly cancellation
condition was seen to translate into a condition on the associated 
affine Lie algebra.  
\\[.3cm]
{\bf Acknowledgments.}
AB thanks Department of Mathematics and Statistics, Concordia University, 
Montr\'eal, for its kind hospitality during the preparation of this work.  
He is very grateful to J. McKay for the invitation, discussions and 
encouragement, and thanks E.H. Saidi for previous collaborations
related to this work. JR thanks F. Lesage for helpful discussions.


\begin{thebibliography}{99}

\bibitem{KV} 
 S. Kachru, C. Vafa, {\em Exact results for $N=2$
 compactifications of heterotic strings}, Nucl. Phys. {\bf B450}
 (1995) 69, hep-th/9505105.

\bibitem{KLMVW} A. Klemm, W. Lerche, P. Mayr, C. Vafa, N. Warner,
 {\em Self-dual strings and $N=2$ supersymmetric field theory},
 Nucl. Phys. {\bf B477} (1996) 746, hep-th/9604034.

\bibitem{KatzVafa} 
 S. Katz, C. Vafa, {\em Matter from geometry},  
 Nucl. Phys. {\bf B497} (1997) 146, hep-th/9606086.

\bibitem{KKV} 
 S. Katz, A. Klemm, C. Vafa, 
 {\em  Geometric engineering of quantum field theories}, 
 Nucl. Phys. {\bf B497} (1997) 173, hep-th/9609239. 

\bibitem{BJPSV} M. Bershadsky, A. Johansen, T. Pantev, V. Sadov, C. Vafa,
 {\em F-theory, geometric engineering and $N=1$ dualities},
 Nucl. Phys. {\bf B505} (1997) 153, hep-th/9612052.

\bibitem{KMV} 
 S. Katz, P. Mayr, C. Vafa, {\em Mirror symmetry and exact
  solution of $4d$  $N=2$ gauge theories I}, Adv. Theor. Math. Phys.
  {\bf 1} (1998) 53, hep-th/9706110.

\bibitem{Mayr} 
 P. Mayr, {\em  Geometric construction of $N=2$ gauge theories}, 
 Fortsch. Phys. {\bf 47} (1999) 39, hep-th/9807096.

\bibitem{BFS} 
 A. Belhaj, A.E. Fallah, E.H. Saidi, {\em On the non-simply
  mirror geometries in type II strings}, Class. Quant. Grav. {\bf 17}
  (2000) 515.

\bibitem{abs}
 M. AitBenhaddou, A. Belhaj, E.H. Saidi, to appear.

\bibitem{CSU}
 M. Cvetic, G. Shiu, A.M. Uranga,  {\em  Three-family supersymmetric 
 standard-like models from intersecting brane worlds},  
 Phys. Rev. Lett. {\bf 87} (2001) 201801, hep-th/0107143;
 {\em  Chiral four-dimensional 
 $N=1$ supersymmetric type IIA orientifolds from intersecting D6-branes},  
 Nucl. Phys. {\bf  B615}  (2001) 3, hep-th/0107166;
 {\em Chiral type II orientifold constructions as M
 theory on $G_2$ holonomy spaces}, hep-th/0111179.

\bibitem{HI}
 A. Hanany, A. Iqbal, {\em Quiver theories from D6-branes via mirror 
 symmetry}, JHEP {\bf 0204} (2002) 009, hep-th/010813.

\bibitem{FHHI}
 B. Feng, A. Hanany, Y.-H. He, A. Iqbal, {\em  Quiver theories, 
 soliton spectra and Picard-Lefschetz transformations}, hep-th/0206152.

\bibitem{FH} 
 S. Franco, A. Hanany, {\em  Geometric dualities in 4d field theories and 
 their 5d interpretation}, hep-th/0207006;
 {\em Toric duality, Seiberg duality and 
 Picard-Lefschetz transformations}, hep-th/0212299.

\bibitem{Uranga} 
 A.M. Uranga, {\em  Chiral four-dimensional string 
 compactifications with intersecting D-branes}, hep-th/0301032.

\bibitem{CIM}
 D. Cremades, L.E. Ibanez, F. Marchesano, {\em Yukawa couplings in 
 intersecting D-brane models}, hep-th/0302105.

\bibitem{He} 
 Y.-H. He, {\em $G_2$ quivers}, JHEP {\bf 0302} (2003) 023,
 hep-th/0210127.

\bibitem{Be}
  A. Belhaj, {\em Comments on M-theory on $ G_2$  manifolds and
 (p,q) webs}, hep-th/0303198.

\bibitem{BS} 
 A. Belhaj, E.H. Saidi, {\em  Toric geometry, enhanced non
  simply laced gauge symmetries in superstrings and F-theory
  compactifications}, hep-th/0012131.

\bibitem{BPS} M. Bershadsky, T. Pantev, V. Sadov, {\em F-theory with
 quantized fluxes}, Adv. Theor. Math. Phys. {\bf 3} (1999) 727,
 hep-th/9805056.

\bibitem{BKMT} P. Berglund, A. Klemm, P. Mayr, S. Theisen,
 {\em On type IIB vacua with varying coupling constant},
 Nucl. Phys. {\bf B558} (1999) 178, hep-th/9805189.

\bibitem{V}
 C. Vafa, {\em On $N=1$ Yang-Mills in four dimensions}, 
 Adv. Theor. Math. Phys. {\bf 2} (1998) 497, hep-th/9801139.

\bibitem{acha}
 B.S. Acharya, {\em  M theory, Joyce orbifolds and super Yang-Mills},  
 Adv. Theor. Math. Phys. {\bf 3} (1999) 227, hep-th/9812205.

\bibitem{FW}
 T. Friedmann, E. Witten, {\em Unification scale, proton decay, and
  manifolds of $G_2$ holonomy}, hep-th/0211269.

\bibitem{GVW} 
 S. Gukov, C. Vafa, E. Witten, {\em CFT's from Calabi-Yau 
 four-folds}, Nucl. Phys. {\bf  B584} (2000) 69, Erratum-ibid. 
 {\bf  B608} (2001) 477, hep-th/9906070. 

\bibitem{Vafa96} C. Vafa, {\em Evidence for F-theory}, Nucl. Phys.
 {\bf B469} (1996) 403, hep-th/9602022.

\bibitem{LV} 
 N.C. Leung, C. Vafa, {\em Branes and toric geometry},
 Adv. Theor. Math. Phys. {\bf 2} (1998) 91, hep-th/9711013.



\end{thebibliography}
\end{document}